\documentclass[12pt]{article}
\usepackage{amssymb} 
\usepackage{amsmath}

\font\tengoth=eufm10 \font\sevengoth=eufm7 \font\fivegoth=eufm5
\newfam\gothfam
\textfont\gothfam=\tengoth \scriptfont\gothfam=\sevengoth
  \scriptscriptfont\gothfam=\fivegoth
  \def\goth{\fam\gothfam}

\def\pd#1#2{\frac{\partial#1}{\partial #2}}
\def\R{{\mathbb R}}

\def\matriz#1#2{\left( \begin{array}{#1} #2 \end{array}\right)}

\def\ad{\mathop{\rm ad}\nolimits}
\def\Ad{\mathop{\rm Ad}\nolimits}

\newtheorem{theorem}{Theorem}

\title{Motion on Lie groups and its applications\\ in Control Theory}
\author{Jos\'e F. Cari\~nena \\ 
Departamento de F\'{\i}sica Te\'orica, Facultad de Ciencias, \\
Universidad de Zaragoza,
50009 Zaragoza, Spain \\ 
e-mail: jfc@posta.unizar.es \\[2ex]
Jes\'us Clemente-Gallardo\\
Instituto de Sistemas y Robotica,\\ Departamento de
 Engenharia Electrotecnica, \\ 
Universidade de Coimbra-Polo II,       \\
Pinhal de Marrocos, 3030-290 Coimbra, Portugal \\
e-mail: jesus@isr.uc.pt  \\[2ex] 
Arturo Ramos \\ 
Dipartimento di Matematica Pura ed Applicata, \\ 
Universit\`a degli Studi di Padova,\\ 
Via G. Belzoni 7, I-35131 Padova, Italy \\
e-mail: aramos@math.unipd.it \\
              \\
}
\date{}
\begin{document}

\maketitle

\vfil\eject

\begin{abstract}

The usefulness in control theory of the geometric theory of motion
on Lie groups and homogeneous spaces will be shown. 
We quickly review some recent results concerning 
two methods to deal with these systems, namely,
a generalization of the method proposed by 
Wei and Norman for linear systems, and a reduction procedure.
This last method allows us to reduce the equation on 
a Lie group $G$ to that on a subgroup $H$, provided  
a particular solution of an associated problem in $G/H$ is known.
These methods are shown to be very appropriate to deal with 
control systems on Lie groups and homogeneous spaces,
through the specific examples of the planar rigid body 
with two oscillators and the front-wheel driven kinematic car.
\end{abstract}


\vfil\eject

\section{Introduction}
Mechanical systems whose configuration space is 
a Lie group are known to be of great relevance both in mechanics and control
theory. The simplest example is 
that of a point particle moving freely on ${\mathbb{R}}^3$, 
for which the configuration space can be identified with 
the Abelian group of translations in a three-dimensional space. 
The motions of the rigid body are described by the Euclidean 
group in three dimensions, $E_3=T_3\odot SO(3,\R)$, 
and for the rigid body with a fixed point the configuration 
space is $SO(3,\R)$. 
However, there are also other systems 
which appear often in control theory which can 
be formulated in homogeneous spaces and reduced to problems of motion on Lie
groups.  The techniques used here have been developed 
in \cite{CGM,CGM01,CarRamGra,CarRamcinc}, 
and firstly applied in control theory in \cite{CarRam02b,CarRam02d}.
The paper is organized as follows. In the next section we review briefly
the geometric theory of motion on Lie groups and homogeneous spaces. 
In Section~3 we recall two basic methods
for dealing with such systems, and in Section~4 we
illustrate the theory through its application on 
two control systems, namely, the planar rigid body 
with two oscillators and the front-wheel driven kinematic car.

\section{Motion on Lie groups and homogeneous spaces}

Given an $r$-dimensional connected Lie group $G$, 
the set of curves $\gamma:{\mathbb R}\to G$, $t\mapsto g(t)$,
can be endowed with a group structure by means of the point-wise 
composition law, $\gamma_2*\gamma_1:t\mapsto g_2(t)\,g_1(t)$, 
for all $t\in{\mathbb R}$. 
Moreover, $G$ acts by right and left translations on itself, and
 if $g(t)$ defines a curve $\gamma$ starting from $g(0)=e$, 
then, for each $g_0\in G$, the maps $\gamma*\gamma_0:t\mapsto g(t)\,g_0$
and $\gamma_0*\gamma:t\mapsto g_0\,g(t)$ are two curves starting from $g_0$,
which are called, respectively, the right and left translation 
 of $\gamma$ by $g_0$. 

In addition, the map $\dot \gamma:{\mathbb R}\to G$ 
given by $t\mapsto \dot g(t)$ defines a vector field along the curve $g(t)$, 
and then, using the right translation by $g^{-1}(t)$, 
we obtain a curve in $T_eG$ which can be written as
\begin{equation}
\dot g(t)\,g^{-1}(t)=-\sum_{\alpha=1}^r b_\alpha(t)a_\alpha\ ,
\label{eqingr}
\end{equation} 
where $\{a_1,\ldots,a_r\}$ is a basis of the 
tangent space $T_eG$ at the neutral element $e\in G$. 
The minus sign on the right hand side is a matter of convention.
The left hand side should be understood as $R_{g^{-1}(t)*g(t)}(\dot g(t))$, 
although for the particular case in which $G\subset GL(n,{\mathbb R})$, 
for certain $n$, it reduces to the above expression.

A first important point is that the curve $g(t)$ can be recovered 
as the only solution of equation (\ref{eqingr}) starting from $e\in G$.  
Any other solution is obtained from the previous $g(t)$ by right translation, 
and, in this sense, (\ref{eqingr}) is right-invariant. 
In other words, $g(t)$ is the integral curve starting 
from $e\in G$ of the $t$-dependent vector field in $G$
\begin{equation} 
X(g,t)=-\sum_{\alpha=1}^r b_\alpha(t)\,X^R_\alpha(g)\ , \label{tdepvfG}
\end{equation}
where $X^R_\alpha$ is the right-invariant vector field in $G$ whose 
value in $e\in G$ is $a_\alpha$, $X_\alpha^R(g)=R_{g*e}(a_\alpha)$. 
Similarly, $X_\alpha^L(g)=L_{g*e}(a_\alpha)$ denotes the left-invariant 
vector field on $G$ determined by $a_\alpha$.
The left-invariant vector fields in $G$ close on a finite-dimensional 
Lie subalgebra $\goth g$ of ${\goth X}(G)$, which is called the Lie 
algebra of $G$. The right-invariant vector fields close on a 
finite-dimensional Lie algebra opposite to $\goth g$.

Let us consider now a transitive action $\Phi:G\times M\to M$
of $G$ on a manifold $M$ (which is then called a homogeneous space of $G$). 
Each choice of a point $x_0\in M$ allows us to identify $M$ 
with a space of left cosets, i.e., $M=G/H$, where $H$ is the stability 
subgroup of the point $x_0$ with respect to $\Phi$. 
Different choices for $x_0$ lead to conjugate subgroups.
We recall that $G$ can be regarded as the total space of the
principal bundle $(G,\tau,G/H)$ over $G/H$, where $\tau:G\mapsto G/H$ 
is the canonical projection.

The important point now is \cite{CarRamGra} 
that the right-invariant vector fields $X^R_\alpha$ 
are $\tau$-projectable, the projections being 
the fundamental vector fields in $M$ associated  
to the natural left action of $G$ on $M=G/H$, identified with $\Phi$.
More explicitly, we have $X_\alpha(x)=X_{a_\alpha}(x)=\Phi_{x*e}(-a_\alpha)$, 
where $x=gH$, and $\tau_{*g}X_\alpha^R(g)=-X_\alpha(gH)$.
Consequently, the vector field (\ref{tdepvfG}) projects onto 
the $t$-dependent vector field in $M$ 
\begin{equation} 
X(x,t)=\sum_{\alpha=1}^rb_\alpha(t)\,X_\alpha(x)\ ,\label{vfLieinM}
\end{equation}
giving rise in this way to a system of differential equations 
\begin{equation}
\dot x^i=\sum_{\alpha=1}^rb_\alpha(t)\,X^i_\alpha(x)\ ,
\qquad i=1,\ldots,n={\rm dim}\,M\, .\label{LieinM}
\end{equation}

This relation between both systems tell us that the solution of 
(\ref{LieinM}) starting from the point $x_0\in M$, 
\emph{arbitrary} but fixed, is given by $x(t)=\Phi(g(t),x_0)$, 
where $g(t)$ is the solution of (\ref{eqingr}) starting 
from the identity. 
Note that the vector fields arising in (\ref{LieinM}) close on the 
Lie algebra $\goth g$. 

Conversely, if a system of type (\ref{LieinM}) is defined by complete 
vector fields closing on a finite dimensional Lie algebra $\goth g$,
then it 
can be regarded as a system of the described type, taking a Lie group 
$G$ such that the vector fields are the fundamental vector fields
of its action on the manifold $M$ given by the integration of their flows. 

In this sense equation (\ref{eqingr}) has a universal character, 
since the knowledge of its solution starting from the identity 
is enough to write the general solution of each associated system of 
type (\ref{LieinM}) on each homogeneous space for $G$. 
Unfortunately, finding the desired particular solution 
of (\ref{eqingr}) may be a difficult problem. 
We briefly recall here two methods for dealing with such equations. 
The first one is based on a generalization \cite{CarMarNas,CarRamcinc}
of the method proposed by Wei and Norman for linear systems \cite{WN,WN2}. 
The second one is a reduction procedure allowing us to 
reduce the problem to a similar one in a subgroup, 
when a solution of the associated  problem in the corresponding 
homogeneous space is known \cite{CGM,CGM01,CarRamGra,CarRamcinc}. 

\section{The generalized Wei--Norman method and the reduction procedure\label{gen_WR_red_proc}}

The main idea of the generalization of the Wei--Norman method 
consists on writing the curve $g(t)$  solution of (\ref{eqingr}) starting from the identity,  
in terms of the second kind canonical coordinates with respect to 
a basis of the Lie algebra $\goth g$, $\{a_1,\,\dots,\,a_r\}$, 
for all $t$, i.e., we write
$
g(t)=\prod_{\alpha=1}^{r}\exp(-v_\alpha(t)a_\alpha)\,.
$
Then, the differential equation (\ref{eqingr}) transforms into
a system of differential equations for the $v_\alpha(t)$, and we have
 to find  the solution determined by 
the initial conditions $v_\alpha(0)=0$ for all $\alpha=1,\,\dots,\,r$.
A simple calculation shows \cite{CarRamcinc} that
\begin{eqnarray}
R_{g(t)^{-1}\,*g(t)}(\dot g(t))
=-\sum_{\alpha=1}^r \dot v_\alpha \left(\prod_{\beta<\alpha} 
\exp(-v_\beta(t) \ad(a_\beta))\right)a_\alpha\,,                \nonumber
\end{eqnarray}
and then, substituting into equation (\ref{eqingr}) 
we obtain the fundamental expression of the Wei--Norman method 
\begin{equation}
\sum_{\alpha=1}^r \dot v_\alpha \left(\prod_{\beta<\alpha} 
\exp(-v_\beta(t) \ad(a_\beta))\right)a_\alpha
=\sum_{\alpha=1}^r b_\alpha(t) a_\alpha\,,
\label{eq_met_WN}
\end{equation}
with $v_\alpha(0)=0$, $\alpha=1,\,\dots,\,r$. The resulting system of
differential equations for the functions $v_\alpha(t)$ is integrable  
by quadratures if the Lie algebra is solvable \cite{WN,WN2}, and in particular, 
for nilpotent Lie algebras. 

On the other hand, given an equation like (\ref{eqingr}) on a 
Lie group $G$, it may happen that the only non-vanishing 
coefficients $b_\alpha(t)$ are those corresponding to a 
subalgebra $\goth h$ of $\goth g$. In that case, the equation 
reduces to a simpler equation on a subgroup, involving less coordinates. 

Now, the important result is that the general situation 
can be reduced to this simpler one under certain conditions.
In fact, we can show that the problem of finding the solution of
(\ref{eqingr}) starting at $e\in G$ can be reduced to the one of 
solving a similar equation in a subgroup $H$, 
provided that one particular solution $\tilde g_1(t)$ for 
the system defined by the left action $\Phi$ of $G$ on the
homogeneous space $M=G/H$ is given. 
The result is as follows \cite{CarRamGra}: 

\begin{theorem}
\label{teor_reduccion}  
Every integral curve of the time-dependent vector 
field (\ref{tdepvfG}) in the group $G$ can be written in the 
form $g(t)=g_1(t)\,h(t)$, where $g_1(t)$ is a curve projecting 
onto a solution $\tilde g_1(t)$ of the system of type (\ref{LieinM})
associated to the left action $\Phi$ on the homogeneous space $G/H$, 
and $h(t)$ is a solution of an equation of type (\ref{eqingr}) 
but for the subgroup $H$, given explicitly by
\begin{equation}
\label{eq:--5}
(\dot h\, h^{-1})(t) 
=-\Ad(g_1^{-1}(t))\left(\sum_{\alpha=1}^r b_\alpha(t)a_\alpha
+(\dot g_1\,g_1^{-1})(t)\right)\in T_eH\ .
\end{equation}
\end{theorem}

This result allows us to understand from a group theoretical point of view, 
as an example, the classical integration theorems for the Riccati equation involving one, 
two or three of its particular solutions \cite{CarRam99}, see also \cite{CarRamGra,CarRamcinc}. 

\section{Some illustrative examples in control theory\label{ex_con_th}}

We will apply the previous techniques to problems appearing 
in the context of control theory, after a very brief description 
of such kind of problems. 

Roughly speaking, a control system is a dynamical system which 
depends on a set of control functions, $u_{\alpha}$, which model
magnitudes to be varied somehow externally to the system,
in order to modify its behaviour according to a specific purpose.
For example, it could be desired to reach one point from another one
of the configuration space, or to minimize some cost functional 
along the evolution of the system with prescribed initial 
and final conditions.

{}From a geometric point of view, we can describe the framework 
of a control system as a bundle $B$ (usually a trivial vector bundle) 
on the state space manifold $M$, with projection $\pi_{B}:B\to M$. The control
dynamical system corresponds then to the integral curves of a vector field
along  the projection $\pi_{B}$, which in local coordinates, $(x^i,u_\alpha)$, 
reads
\begin{equation}
\label{eq:-}
\dot x^i=X^i(x,u)\,, \quad x\in M\,,\quad u\in B_{x}=\pi_{B}^{-1}(x)\ .
\end{equation} 
A special class is those of the control systems affine in the controls:
\begin{equation}
\label{eq:--2}
\dot x^i=X_0^i(x)+\sum_{\alpha=1}^{r}u_{\alpha}\,X^i_{\alpha}(x)\ ,
\qquad i=1,\ldots,n={\rm dim}\,M\ .
\end{equation}
The vector field $X_0^i\,\partial/\partial x^i\in {\goth{X}}(M)$ is
called the \emph{drift}\/ of the system. In principle, the control 
functions $u_{\alpha}$ are supposed to depend on time 
(in the control theory terminology the system 
is then operating in \emph{open loop}). 

We are now interested in those which are drift-less, i.e., $X_0=0$,
since they become a system of type (\ref{LieinM}) as soon as the set of
vector fields $\{X_{\alpha}^i\partial/\partial x^i\}$ closes on  
a finite-dimensional Lie algebra. 

It may also happen that starting with a control system that 
is of the form (\ref{LieinM}), but where the vector fields 
$\{X_{\alpha}^i\partial/\partial x^i\}$ do not close on a 
Lie algebra, an appropriate \emph{feedback transformation} 
$u_\alpha(x,t)=\sum_{\beta=1}^r f_{\alpha\beta}(x)v_\beta(t)$ 
could lead to a new system written as
\begin{equation}
\dot x^i=\sum_{\alpha=1}^r v_\alpha(t)\,Y^i_\alpha(x)\ ,
\qquad i=1,\ldots,n={\rm dim}\,M\, ,
\end{equation}
where $Y_\alpha(x)=\sum_{i=1}^r f_{\beta\alpha}(x)X_\beta(x)$
do close a finite-dimensional Lie algebra.

We will illustrate by means of simple examples how systems 
of type (\ref{LieinM}), both directly or as a consequence 
of a feedback transformation, appear in control theory. 
In addition, with the general theory developed in previous
sections, we will show how it is possible to solve, reduce,
or relate systems of this kind formulated on different 
homogeneous spaces. 

\subsection{Planar rigid body with two oscillators\label{planar_rb_two_osc}}

This example comes {}from the consideration of the 
optimal control problem of a planar rigid body with 
two oscillators \cite{YanKriDay96}. 
The control system of interest has the state space $\R^2\times S^1$, 
with coordinates $(x_1,\,x_2,\,\theta)$ 
\begin{equation}
\dot x_1=b_1(t)\,,\quad\dot x_2=b_2(t)\,,\quad\dot \theta=x_1^2 b_2(t)-x_2^2 b_1(t)\,,
\label{syst_plan_rb_two_oscil}
\end{equation}
where $b_1(t)$ and $b_2(t)$ are the control functions.  
This system is similar to the celebrated Brockett
nonholonomic integrator system \cite{Bro82}, but where the third equation 
is quadratic in the coordinates instead of linear, and the meaning
of the third coordinate is now an angle.  

The solutions of the system (\ref{syst_plan_rb_two_oscil}) are 
the integral curves of the time-dependent vector field $b_1(t)\, X_1+b_2(t)\, X_2$, 
with 
\begin{equation}
X_1=\pd{}{x_1}-x_2^2\pd{}{\theta}\,,\quad\quad X_2=\pd{}{x_2}+x_1^2\pd{}{\theta}\,.
\label{vf_1st_real_pl_rb_2osc}
\end{equation}
The Lie brackets 
$$
X_3=[X_1,\,X_2]=2(x_1+x_2)\pd{}{\theta}\,,\quad X_4=[X_1,\,X_3]=2\pd{}{\theta}\,,
$$ 
jointly with $X_1,\,X_2$, make up a linearly independent 
set in points with $x_1\neq -x_2$, and the set $\{X_1,\,X_2,\,X_4\}$ 
spans the tangent space at every point of $\R^2\times S^1$. 
According to Chow's theorem \cite{Cho4041}, every two such points can
be joined by appropriate piecewise constant controls $b_1(t)$ and $b_2(t)$, 
therefore the system is controllable.
In addition, the set $\{X_1,\,X_2,\,X_3,\,X_4\}$ closes on the 
nilpotent Lie algebra defined by the non-vanishing Lie brackets 
\begin{eqnarray}
&& [X_1,\,X_2]=X_3\,,\quad\quad [X_1,\,X_3]=X_4\,, \quad\quad[X_2,\,X_3]=X_4\,,
\label{comm_camp_vec_rb_two_osc}
\end{eqnarray}
isomorphic to a nilpotent Lie algebra, denoted as $\goth g_4$, 
which can be regarded as a central extension of the Heisenberg Lie 
algebra ${\goth h}(3)$ by $\R$.  In fact, taking the 
basis $\{a_1,\,a_2,\,a_3,\,a_4\}$ of $\goth g_4$ for which the non-vanishing 
Lie products are 
\begin{eqnarray}
&& [a_1,\,a_2]=a_3\,,\quad\quad [a_1,\,a_3]=a_4\,, \quad\quad[a_2,\,a_3]=a_4\,,
\label{comm_lie_alg_rb_two_osc}
\end{eqnarray}
then the center $\goth z$ of the algebra is generated by $\{a_4\}$, and 
the factor Lie algebra ${\goth g_4}/{\goth z}$ is isomorphic to the 
Heisenberg Lie algebra ${\goth h}(3)$.

Let $G_4$ be the connected and simply connected nilpotent 
Lie group such that its Lie algebra is the previous $\goth g_4$.
The right-invariant system of type (\ref{eqingr}) on $G_4$ 
corresponding to (\ref{syst_plan_rb_two_oscil}) is
\begin{equation}
R_{g(t)^{-1}*g(t)}(\dot g(t))=-b_1(t)a_1-b_2(t)a_2\,.
\label{eq_grup_gr_rb_two_osc_b3b4_nulo}
\end{equation}
Let us solve it by the Wei--Norman method.
We write the solution of (\ref{eq_grup_gr_rb_two_osc_b3b4_nulo})
starting from the identity as the product of exponentials
\begin{equation}
g(t)=\exp(-v_1(t)a_1)\exp(-v_2(t)a_2)\exp(-v_3(t)a_3)\exp(-v_4(t)a_4)\,,
\label{fact_WN_1_rb_two_oscil}
\end{equation}
and using  the expression  
of the adjoint representation of $\goth g_4$, and its exponentiation,
and applying then (\ref{eq_met_WN}), we find the system
\begin{eqnarray}
\dot v_1=b_1(t)\,,
\quad\dot v_2=b_2(t)\,,
\quad\dot v_3=b_2(t)\,v_1\,,
\quad\dot v_4=b_2(t)\,v_1(v_1/2+v_2)\,,
\label{sist_vs_rb_two_oscil}
\end{eqnarray}
with initial conditions $v_1(0)=v_2(0)=v_3(0)=v_4(0)=0$. 
The solution is easily found by quadratures: 
if we denote $B_i(t)=\int_0^t b_i(s)\,ds$, $i=1,\,2$, then,
\begin{eqnarray}
&&v_1(t)=B_1(t)\,,
\quad v_2(t)=B_2(t)\,,
\quad v_3(t)=\int_0^t b_2(s)B_1(s)\,ds\,.                       \nonumber\\
&&v_4(t)=\int_0^t b_2(s)\left(\frac 1 2 B_1^2(s)+B_1(s)B_2(s)\right)\,ds\,.
\label{sol_vs__rb_two_oscil}
\end{eqnarray}
Other orderings in the factorization (\ref{fact_WN_1_rb_two_oscil}) 
are possible, with similar results.

Now, we can find the expressions of the action $\Phi$ of $G_4$ 
on the configuration manifold $\R^2\times S^1$ such that $X_i$ 
be the infinitesimal generator associated to $a_i$ for 
each $i\in\{1,\,\dots,\,4\}$, and of the composition law of $G_4$.
We will use canonical coordinates of the second kind in $G_4$ defined 
by $g=\exp(a a_1)\exp(b a_2)\exp(c a_3)\exp(d a_4)$, and then,
integrating and composing accordingly the flows of 
the vector fields $X_i$, the action reads
$\Phi:G_4\times(\R^2\times S^1)\to\R^2\times S^1$,
\begin{eqnarray}
&&\Phi((a,\,b,\,c,\,d),\,(x_1,\,x_2,\,\theta))          
=(x_1-a,\,x_2-b,                                        \nonumber\\
&&\quad\quad\quad\quad\quad\,\theta+a x_2^2-b x_1^2-2(a b+c)x_2-2 c x_1+a b^2-2 d)\,,
\label{accion_rb_two_oscil_coord_can_2nd_class}
\end{eqnarray}
and the composition law of $G_4$ reads
\begin{eqnarray}
&&(a,\,b,\,c,\,d)(a^\prime,\,b^\prime,\,c^\prime,\,d^\prime)
=(a+a^\prime,\,b+b^\prime,\,c+c^\prime-b a^\prime,\,    \nonumber\\     
&&\quad\quad\quad\quad\quad\quad\quad\quad\quad 
d+d^\prime-c(a^\prime+b^\prime)+b a^\prime (b+2 b^\prime+a^\prime)/2)\,.
\nonumber 
\end{eqnarray}
The neutral element is represented by $(0,\,0,\,0,\,0)$ in these coordinates. 

The general solution of (\ref{syst_plan_rb_two_oscil}) can be calculated
by means of the solution of the Wei--Norman system (\ref{sist_vs_rb_two_oscil}) as
\begin{eqnarray}
&&\Phi((-v_1,\,-v_2,\,-v_3,\,-v_4),\,(x_{10},\,x_{20},\,\theta_0))
=(x_{10}+v_1,\,x_{20}+v_2,                                      \nonumber\\
&&\quad\quad\quad\quad\quad\,\theta_0-v_1 x_{20}^2+v_2 x_{10}^2-2(v_1 v_2-v_3)x_{20}
+2 v_3 x_{10}-v_1 v_2^2+2 v_4)\,,                               \nonumber
\end{eqnarray}
where $v_1=v_1(t)$, $v_2=v_2(t)$, $v_3=v_3(t)$ and $v_4=v_4(t)$ are given 
by (\ref{sol_vs__rb_two_oscil}), $(x_{10},\,x_{20},\,\theta_0)\in\R^2\times S^1$ 
are the initial conditions and $\Phi$ is 
given by (\ref{accion_rb_two_oscil_coord_can_2nd_class}).

We will show the way the previous reduction theorem applies to
the control system (\ref{syst_plan_rb_two_oscil}). Several 
possibilities of reduction exist, as many as (non-equivalent) 
subgroups of $G_4$. Specifically, we will show how the system
(\ref{eq_grup_gr_rb_two_osc_b3b4_nulo}) can be reduced to 
a control system of Brockett type plus a system on the real line,
performing the reduction with respect to the center of $G_4$.

For a better illustration we parametrize now $G_4$ by canonical 
coordinates of first kind defined by $g=\exp(a a_1+b a_2+c a_3+d a_4)$, 
being then the composition law  
\begin{eqnarray}
&&(a,\,b,\,c,\,d)(a^\prime,\,b^\prime,\,c^\prime,\,d^\prime)
=(a+a^\prime,\,b+b^\prime,\,c+c^\prime+(a b^\prime-b a^\prime)/2,\,     \nonumber\\     
&&\quad d+d^\prime+(a c^\prime-c a^\prime)/2+(b c^\prime-c b^\prime)/2
+(a b^\prime-b a^\prime)(a-a^\prime+b-b^\prime)/12)\,.
\nonumber 
\end{eqnarray}
Thus, the adjoint representation of the group reads 
\begin{equation}
\Ad(a,\,b,\,c,\,d)=\matriz{cccc}{1&0&0&0\\0&1&0&0\\-b&a&1&0
\\-\frac b 2(a+b)-c&\frac a 2(a+b)-c&a+b&1}\,.
\label{Adjoint_G_rb_two_oscil}
\end{equation}
If $g(t)=(a(t),\,b(t),\,c(t),\,d(t))$ is a curve in $G_4$
expressed in the previous coordinates, we obtain
\begin{eqnarray}
R_{g^{-1}*g}(\dot g)
=\matriz{c}{\dot a\\ \dot b \\ \dot c-\frac 1 2(b \dot a-a \dot b) \\ 
\dot d-\frac 1 6(a b+b^2+3 c)\dot a
+\frac 1 6(a^2+a b-3 c)\dot b+\frac 1 2(a+b)\dot c}\,. \nonumber
\end{eqnarray}

To perform the reduction we select the subgroup $H$ 
to be the center of $G_4$, which is generated by $a_4$. 
The relevant factorization 
is $(a,\,b,\,c,\,d)=(a,\,b,\,c,\,0)(0,\,0,\,0,\,d)$. 
Therefore, we can describe the homogeneous space $M=G_4/H$ 
by means of the projection
\begin{eqnarray*}
\tau:G_4&\longrightarrow&G_4/H  \\
(a,\,b,\,c,\,d) &\longmapsto& (a,\,b,\,c)\,,
\end{eqnarray*}
associated to the previous factorization.
We take coordinates $(y_1,\,y_2,\,y_3)$ in $M$.
Thus, the left action of $G_4$ on such a homogeneous space reads
\begin{eqnarray*}
\Phi:G_4\times M&\longrightarrow& M                             \\
((a,\,b,\,c,\,d),\,(y_1,\,y_2,\,y_3))&\longmapsto&
(y_1+a,\,y_2+b,\,y_3+c+(a y_2-b y_1)/2)\,.
\end{eqnarray*}
The corresponding fundamental vector fields are
$$
X_1^H=-\pd{}{y_1}-\frac{y_2}{2}\pd{}{y_3}\,,
\quad X_2^H=-\pd{}{y_2}+\frac{y_1}{2}\pd{}{y_3}\,,
\quad X_3^H=-\pd{}{y_3}\,,\quad X_4^H=0\,,
$$
which span the tangent space at each point of $M$, and, in addition, 
satisfy the Lie brackets $[X_1^H,\,X_2^H]=X_3^H$, $[X_1^H,\,X_3^H]=X_4^H=0$ 
and $[X_2^H,\,X_3^H]=X_4^H=0$. 

Now, let the desired solution 
of (\ref{eq_grup_gr_rb_two_osc_b3b4_nulo}) be factorized  as the product
$$
g_1(t)h(t)=(y_1(t),\,y_2(t),\,y_3(t),\,0)(0,\,0,\,0,\,d(t))\,,
$$
where $g_1(t)$ projects onto the solution 
$\tau(g_1(t))=(y_1(t),\,y_2(t),\,y_3(t))$, 
with initial conditions $(0,\,0,\,0)$, 
of the Lie system on the homogeneous space $M$ 
associated to (\ref{eq_grup_gr_rb_two_osc_b3b4_nulo}),
\begin{equation}
\dot y_1=-b_1(t)\,,\quad \dot y_2=-b_2(t)\,,
\quad\dot y_3=\frac 1 2 (b_2(t) y_1-b_1(t) y_2)\,.
\label{Lie_sys_H3_hom_sp1}
\end{equation}
Then, we reduce the problem to a Lie system in the subgroup $H$
for $h(t)=(0,\,0,\,0,\,d(t))$, with $h(0)=e$, i.e., $d(0)=0$. 
The expression of this last system is given by 
Theorem~\ref{teor_reduccion}, i.e.,
$$
\dot d=\frac{1}{12}((y_1+y_2)(b_1 y_2-b_2 y_1)-6 y_3(b_1+b_2))\,,
$$
which is a system on $H\cong\R$, solvable by one quadrature.
The system (\ref{Lie_sys_H3_hom_sp1}) is of Brockett 
type \cite{Bro82} (indeed they are related by the simple change of 
coordinates $x=-y_1$, $y=-y_2$ and $z=-2 y_3$), 
and therefore we obtain two interesting results. 
Firstly, that solving a system of type (\ref{syst_plan_rb_two_oscil}) 
can be reduced to solving first a system of Brockett type and then 
to solving a Lie system in $\R$, which is immediate.
Secondly, that the Brockett system can be regarded as a Lie system on $H(3)$ 
written, moreover, in terms of canonical coordinates of first kind of such a group.

As an interesting open problem, it remains to investigate the
interrelations the corresponding optimal control problems might have 
with respect to this reduction.
 
\subsection{Front-wheel driven kinematic car\label{Kin_car}}

The front-wheel driven kinematic car 
has been considered by a number of authors, mainly with regard 
to the nonholonomic motion planning problem, and as such is made 
nilpotent by a state space feedback transformation 
\cite{LafSus91,LafSus93,Mur93,MurSas91,MurSas93}. It is 
a simple model of a car with front and rear wheels. 
The distance between the rear and front axles is $l$, 
which we will take as 1 for simplicity.

The configuration of the car is determined by the Cartesian 
coordinates $(x,\,y)$ of the rear wheels, the angle of the 
car body $\theta$ with respect to the horizontal coordinate axis, 
and the steering front wheel angle $\phi\in I=(-\pi/2,\,\pi/2)$ relative 
to the car body. The configuration space is therefore $\R^2\times S^1\times I$, 
with coordinates $(x,\,y,\,\theta,\,\phi)$.
The controls of the system are the velocity of the rear (or sometimes front) 
wheels and the turning speed of the front wheels.
For a schematic picture of the system, see, e.g., \cite{Mur93,MurSas93}. 

The control system for this model can be written as \cite{Mur93,MurSas93} 
(compare with \cite[Eq. (13.7)]{LafSus91})
\begin{equation} 
\dot x=c_1(t)\,,
\quad \dot y=c_1(t)\tan \theta \,,
\quad \dot \phi=c_2(t)\,,
\quad \dot \theta=c_1(t)\tan \phi\sec\theta\,.
\label{sist_Murr_Sas}
\end{equation} 
Note that this system is defined for 
angles $\theta$ with $\cos\theta\neq 0$. 
We therefore restrict $\theta\in I$ as well. 
The solutions of (\ref{sist_Murr_Sas}) are the integral 
curves of the time-dependent vector field $c_1(t) Y_1+c_2(t) Y_2$, where 
\begin{equation}
Y_1=\pd{}{x}+\tan\theta\pd{}{y}+\tan\phi\,\sec\theta\pd{}{\theta}\,,
\quad Y_2=\pd{}{\phi}\,.
\label{inp_vf_kin_car}
\end{equation}
Taking the Lie brackets 
\begin{eqnarray*}
&&Y_3=[Y_1,\,Y_2]=-\sec\theta \sec^2\phi\pd{}{\theta}\,,
\quad Y_4=[Y_1,\,Y_3]=\sec^2\theta \sec^2\phi\pd{}{y}\,,
\end{eqnarray*}
we see that $\{Y_1,\,Y_2,\,Y_3,\,Y_4\}$ generate the full tangent space 
at points of the (restricted) configuration space $\R^2\times I\times I$,
thus the system is controllable there. However, (\ref{sist_Murr_Sas}) 
is not a system of type (\ref{LieinM}), since the iterated Lie brackets 
$$
\left[Y_2,\,\left[Y_2,\,\dots \left[Y_2,\,Y_1\right] \cdots \right]\right]
\quad\mbox{or}
\quad\left[Y_1,\,\left[Y_1,\,\dots \left[Y_1,\,Y_2\right] \cdots \right]\right]
$$
generate at each step vector fields linearly independent from 
those obtained at the previous stage, therefore they do not close a 
finite-dimensional Lie algebra.

Notwithstanding, it can be transformed into a nilpotent 
system of type (\ref{LieinM}). In \cite{LafSus91,Mur93,MurSas93} 
it is proposed the following state space feedback transformation 
(it seems that in \cite{Mur93,MurSas93} there are some minor misprints, 
for their expressions do not do the work)
\begin{eqnarray}
c_1(t)=b_1(t)\,\quad c_2(t)=-3 \sin^2\phi \sec^2\theta \sin\theta\,b_1(t)
+\cos^3\theta \cos^2\phi\,b_2(t)\,,
\label{feedback_Kinem_car}
\end{eqnarray}
and then the change of coordinates
\begin{equation}
x_1=x\,,
\quad x_2=\sec^3\theta\tan\phi\,,
\quad x_3=\tan\theta\,,
\quad x_4=y\,,                  
\label{chan_coor_Kinem_car}
\end{equation}
with inverse 
$$
x=x_1\,,
\quad y=x_4\,,
\quad \theta=\arctan x_3\,,
\quad \phi=-\arctan\left(\frac{x_2}{(1+x_3^2)^{3/2}}\right)\,,                  
$$
which transforms (\ref{sist_Murr_Sas}) into the control system
in $\R^4$ with coordinates $(x_1,\,x_2,\,x_3,\,x_4)$ given by 
\begin{equation}
\dot x_1=b_1(t)\,,
\quad \dot x_2=b_2(t)\,,
\quad \dot x_3=b_1(t)\,x_2\,,
\quad \dot x_4=b_1(t)\,x_3\,,
\label{chain_form}
\end{equation}
where the control functions are now $b_1(t)$ and $b_2(t)$.
In other words, we take the new input vector fields
$$
X_1=Y_1-3 \sin^2\phi\sec^2\theta\sin\theta\,\,Y_2\,,\quad               
X_2=\cos^3\theta \cos^2\phi\,\, Y_2\,.
$$

The system (\ref{chain_form}) is usually said to be in \emph{chained form}, 
see \cite{Mur93,MurSas93}. Let us show that it is a system of type (\ref{LieinM}). 
The new input vector fields read, in the coordinates $(x_1,\,x_2,\,x_3,\,x_4)$, as
\begin{equation}
X_1=\pd{}{x_1}+x_2\pd{}{x_3}+x_3 \pd{}{x_4}\,,
\quad X_2=\pd{}{x_2}\,.
\label{inp_vf_kin_car_with_feedb_new_cor}
\end{equation}
The Lie brackets 
$$
X_3=[X_1,X_2]=-\pd{}{x_3}\,,\quad X_4=[X_1,X_3]=\pd{}{x_4}\,,
$$ 
are linearly independent from $X_1$ and $X_2$, and 
$\{X_1,\,X_2,\,X_3,\,X_4\}$ generate the full tangent space 
at every point of the configuration space $\R^4$, so the system
is controllable. On the other hand, the same set closes on the
nilpotent Lie algebra defined by the non-vanishing Lie brackets
\begin{equation}
[X_1,\,X_2]=X_3\,,\quad\quad [X_1,\,X_3]=X_4\,.
\label{comm_rels_Murr_Sas_lin}
\end{equation}
This Lie algebra is isomorphic to a four dimensional 
nilpotent Lie algebra, denoted by $\bar{\goth g}_4$,
which is also a central extension of the Lie algebra ${\goth h}(3)$ by $\R$. 
Indeed, taking the basis $\{a_1,\,a_2,\,a_3,\,a_4\}$ of $\bar{\goth g}_4$ 
such that the non-vanishing defining relations are
\begin{eqnarray}
&& [a_1,\,a_2]=a_3\,,\quad\quad [a_1,\,a_3]=a_4\,,
\label{comm_lie_alg_kin_car_feed}
\end{eqnarray}
then the center $\goth z$ of the algebra is generated by $\{a_4\}$, and 
the factor Lie algebra $\bar{\goth g}_4/{\goth z}$ is isomorphic to ${\goth
  h}(3)$. However, this extension is not equivalent to the extension appearing
in the case of the planar rigid body with two oscillators, 
compare (\ref{comm_lie_alg_kin_car_feed}) with (\ref{comm_lie_alg_rb_two_osc}).

We give now briefly some results concerning the solution of (\ref{chain_form})
by the Wei--Norman method and its reduction to a system of Brockett type plus
a system in $\R$. The calculations are similar to that of the previous
subsection. 

Let $\bar G_4$ be the connected and simply connected nilpotent Lie group 
whose Lie algebra is $\bar{\goth g}_4$. The right-invariant system of type 
(\ref{eqingr}) on $\bar G_4$ corresponding to (\ref{chain_form})
reads as (\ref{eq_grup_gr_rb_two_osc_b3b4_nulo}) but where now
$g(t)$ is the solution curve in $\bar G_4$ starting from the identity, 
and $\{a_1,\,a_2,\,a_3,\,a_4\}$ is the previous basis of $\bar{\goth g}_4$.
Writing $g(t)=\exp(-v_1 a_1)\exp(-v_2 a_2)\exp(-v_3 a_3)\exp(-v_4 a_4)$,
the Wei--Norman method gives the system
\begin{equation}
\dot v_1=b_1(t)\,,
\quad\dot v_2=b_2(t)\,,
\quad\dot v_3=b_2(t) v_1\,,    
\quad\dot v_4=b_2(t)\,\frac{v_1^2}{2}\,,
\label{sist_vs_Murr_Sas}
\end{equation}
with initial conditions $v_1(0)=v_2(0)=v_3(0)=v_4(0)=0$, which is 
easily integrable by quadratures. The action $\Phi$ of $\bar G_4$ on $\R^4$ 
corresponding to the infinitesimal generators $\{X_i\}$ reads, 
parametrizing $g\in \bar G_4$ by $g=\exp(a a_1)\exp(b a_2)\exp(c a_3)\exp(d
a_4)$,  
\begin{eqnarray*}
\Phi:\bar G_4\times\R^4&\longrightarrow&\R^4                    \\
((a,\,b,\,c,\,d),\,(x_1,\,x_2,\,x_3,\,x_4))
&\longmapsto&(\bar x_1,\,\bar x_2,\,\bar x_3,\,\bar x_4)\,,
\end{eqnarray*}
where $\bar x_1=x_1-a$, $\bar x_2=x_2-b$, $\bar x_3=x_3-a x_2+a b+c$ 
and $\bar x_4=x_4-a x_3+a^2 x_2/2-a^2 b/2-a c-d$.
The composition law reads
$$
(a,\,b,\,c,\,d)(a^\prime,\,b^\prime,\,c^\prime,\,d^\prime)
=(a+a^\prime,\,b+b^\prime,\,c+c^\prime-b a^\prime,\,    
d+d^\prime-ca^\prime+b a^{\prime\,2}/2)\,,
\label{group_law_kin_car_feed_2nd_can_cord}
$$
the neutral element being represented by $(0,\,0,\,0,\,0)$.

Thus, the general solution of (\ref{chain_form}) is 
\begin{eqnarray}
&&\Phi((-v_1,\,-v_2,\,-v_3,\,-v_4),\,(x_{10},\,x_{20},\,x_{30},\,x_{40}))
=(x_{10}+v_1,\,x_{20}+v_2,\,                                    \nonumber\\
&&\quad\,x_{30}+v_1 x_{20}+v_1 v_2-v_3,\,
x_{40}+v_1 x_{30}+v_1^2 x_{20}/2+v_1^2 v_2/2-v_1 v_3+v_4)\,,    \nonumber
\end{eqnarray}
where $v_1=v_1(t)$, $v_2=v_2(t)$, $v_3=v_3(t)$ and $v_4=v_4(t)$ are 
the solution of (\ref{sist_vs_Murr_Sas}) and the initial conditions are
$(x_{10},\,x_{20},\,x_{30},\,x_{40})\in\R^4$.

Due to the Lie algebra structure of $\bar{\goth g}_4$, by quotienting by 
the center, we can reduce the solution of the equation in the group 
$\bar G_4$ (and hence of (\ref{chain_form})) to two other problems: 
one, a system in $H(3)$ which is of Brockett type, 
and then we have to integrate a system in $\R$. 
Taking canonical coordinates of first kind, and following analogous 
steps to that of the previous subsection, 
we have to solve first (\ref{Lie_sys_H3_hom_sp1}), and then 
\begin{equation}
\dot d=\frac{b_1(t)}2\left(\frac{1}{6}y_1(t)y_2(t)-y_3(t)\right)
-\frac{1}{12}b_2(t)y_1^2(t)\,,          \label{Lie_syst_ideal_red_kin_car_feed}
\end{equation}
which is integrable by one quadrature.

\section*{Acknowledgements}

Partial support of the Spanish DGI, 
project BFM2000-1066-C03-01, is acknowledged.
A.R. has been partially supported by the Spanish Ministerio de Ciencia
y Tecnolog\'{\i}a through a FPI grant and by the European Commission 
funding for the Human Potential Research Network \lq\lq Mechanics and
Symmetry in Europe\rq\rq\ (MASIE), contract HPRN-CT-2000-00113. 
J.C.-G. has been partially supported by the European Union through 
the TMR Network in Nonlinear Control (Contract ERB FMRXCT-970137).

\vfil\eject

\end{document}